\documentclass[aps,pre,notitlepage]{revtex4-1}
\usepackage{graphicx}
\usepackage{subfigure}
\usepackage{amsmath}
\usepackage{amsfonts}
\usepackage{hyperref}
\usepackage{color}

\newcommand{\mutilde}{\widetilde{\mu}}
\newcommand{\muhat}{\widehat{\mu}}
\newcommand{\ka}{\kappa a}
\renewcommand{\r}{\mathbf{r}}

\newcommand{\marked}[1]{{#1}}

\begin{document}
\title{Free energy of cylindrical polyions: analytical results}

\author{Gabriel T\'ellez}
\affiliation{Departamento de F\'{\i}sica, Universidad de los Andes,
  Bogot\'a, Colombia}
\email{gtellez@uniandes.edu.co (corresponding author)}

\author{Emmanuel Trizac}
\affiliation{Laboratoire de Physique Th\'eorique et Mod\`{e}les
  Statistiques ({\it UMR CNRS 8626}),
 Universit\'e Paris-Sud, Universit\'e Paris Saclay, F-91405
 Orsay, France}
\email{trizac@lptms.u-psud.fr}

%\date{\today}

\begin{abstract}
Within the Poisson--Boltzmann (PB) framework useful for a wealth of
charged soft matter problems, we work out the Coulombic grand
potential of a long cylindrical charged polyion in a binary
electrolyte solution of arbitrary valency and for low salt
concentration.  We obtain the exact analytical low-salt asymptotic
expression for the grand potential, derived from known properties of
the exact solutions to the cylindrical PB equation. These results are
relevant for understanding nucleic acid processes. In practice, our
expressions are accurate for arbitrary polyion charges, provided their
radius is smaller than the Debye length defined by the electrolyte.
\end{abstract}

\maketitle

\section{Introduction}

Coulombic interactions play an important role in the physico-chemical
and thermodynamic properties of highly charged polyions in a solution
with added salt \cite{Levin02,Andelman06,Messina09}. In this work, we
concentrate on free energy calculations pertaining to stiff charged
polymers, essential for a number of applications among which melting
and binding of biopolymers such as nucleic acids.  We provide
{analytical} expressions for the free energy of the formation of the
electrical double layer around long cylindrical polyions such as
deoxyribonucleic acid (DNA), \marked{in the infinite dilution limit}. On the
mean field level, where correlations are neglected, the nonlinear
Poisson--Boltzmann (PB) theory provides accurate predictions for most
applications~\cite{VO48, CM83,SD89,SBR06}, which can be tested against
results from Monte-Carlo and molecular dynamics
simulations~\cite{BZ84,MAR89}. Nowadays, with available computer
numerical libraries and programs, it is fairly straightforward to
solve numerically PB equation to obtain many quantities of interest,
including the free energy, for example see~\cite{TBAG03} and
appendix~\ref{app:numerics} {below}. However, it is desirable to
obtain analytic results that give more insight into the dependency of
the free energy on the parameters of the system: the linear charge of
the polyion, the salt concentration and ions valencies. For high salt
concentration situations, an expansion in small curvature around the
planar double layer result can be build to provide results for the
free energy~\cite{S10}. Here, we will concentrate on the opposite
regime of low salt concentration, making use of the known analytic
asymptotic expansion of the solution of PB equation~\cite{CTW77, TW97,
  TW98, TT06, TTexactPB06, TT07}. Due to the chemical equilibrium with
the reservoir, the appropriate ensemble is the grand-canonical
one. Therefore we will concentrate on evaluating the grand
potential. The free energy can be obtained by the usual Legendre
transformation, see appendix~\ref{app:free-energy}.

This work is organized as follows. In Sec.~\ref{sec:general-omega},
after recalling the PB framework and previous results, we derive the
exact low-salt concentration asymptotic analytic expression for the
grand potential. The result is valid for any value of the linear
charge of the polyion. In Sec.~\ref{sec:xi-close-xic}, we provide a
simplification of the general result that is valid for moderate to
highly charged polyions. This expression has the advantage to be valid
for any electrolyte valencies and not limited to 1:1. Finally in
Sec.~\ref{sec:results}, we benchmark our analytic expressions
against numerical evaluation of the grand potential and we discuss our
predictions dependence on the polyion charge, on the salt
concentration and on the electrolyte valencies.

\section{General expression for the grand potential}
\label{sec:general-omega}

Our framework is the nonlinear PB equation \cite{Levin02,Andelman06,Messina09} to describe a cylindrical
polyion in an {infinite} electrolyte medium with dielectric permittivity
$\epsilon$. The persistence length of the polyion is supposed to be
much larger than all other physical lengths of interest, therefore the
polyion is modeled as an infinite cylinder of radius $a$ with uniform
linear charge density $\lambda=-e/b<0$, with $e>0$ the elementary
charge and $b$ the longitudinal distance per unit charge. The system
is in thermal and chemical equilibrium with a salt reservoir at
temperature $T=1/(k_B \beta)$ ($k_B$ is Boltzmann constant) and
chemical potentials $\mu_{\pm}=k_B T \ln(n_{\pm}^0\Lambda_{\pm}^3)$,
where $n_{\pm}^0$ are the ionic bulk densities and $\Lambda_{\pm}$ are
the de Broglie thermal wavelengths of the ions. The electrolyte
valencies are $z_{-}$:$z_{+}$. Both numbers are taken positive with the
convention of writing first the coion valency (here $z_{-}$) then the
counterion valency ($z_{+}$). The charge density of the polyion can be
characterized by the dimensionless parameter $\xi=l_B/b=-\lambda
l_B/e>0$, with $l_B=\beta e^2/\epsilon$ the Bjerrum length
(around 0.71 nm for water at room temperature). The solvent (water) is
modeled as a continuous medium of dielectric relative permitivitty
$\epsilon$. The Debye length $\kappa^{-1}$ is defined by
$\kappa^2=4\pi l_B (z_{+}^2 n_{+}^0 + z_{-}^2 n_{-}^0)$. The
dimensionless electrostatic potential at a radial distance $r$ from
the polyion, $\phi(r)=\beta e y(r)$ (with $y(r)$ the electrostatic
potential), satisfies PB equation
\begin{equation}
  \label{eq:PB}
  \frac{1}{r}\frac{d}{dr}\left(r \frac{d\phi}{dr} \right)
  = \frac{\kappa^2}{z_{+}+z_{-}}\left(e^{z_{-}\phi(r)}-e^{-z_{+}\phi(r)}\right)
  \,,
\end{equation}
with boundary conditions $a\phi'(a)=2\xi$ (Gauss law at contact {with}
the polyion) and $\lim_{r\to\infty} r \phi'(r)=0$ (electroneutrality
of the {system} in the infinite dilution limit considered here). The
grand potential of the system can be obtained by using one of several
{charging processes} as recalled in
appendix~\ref{app:free-energy}~\marked{\cite{M55, TH96,TH97}}. We use
here a {charging} process where $\xi$ is varied from $0$ to its final
value. The dimensionless excess grand potential per unit charge is
$g=\omega/\xi$, with the dimensionless linear density of grand
potential (see definitions~(\ref{eq:def-omega}) and (\ref{eq:def-g})
in appendix~\ref{app:free-energy})
\begin{equation}
  \label{eq:charging}
  \omega = - \int_0^{\xi} \phi_0\, d\xi' \,.
\end{equation}
Here $\phi_0$ is the contact electrostatic potential $\phi_0=\phi(a)$
seen as a function of the charge $\xi'$.

In this work we consider the low salt density regime when $\kappa a\ll
1$. As explained in appendix D of ref.~\cite{TT07}, the short distance
behavior of the electric potential can be obtained by injecting into
PB Eq.~(\ref{eq:PB}) the $o(1)$ approximation $\phi(r)=-2A\ln(\kappa r)+
\ln B + o(1)$ to compute higher order terms of powers $r$. $A$ and $B$
are constants of integration. Summing up all terms of order
$r^{2n(1+z_{+}A)}$ ($n\in\mathbb{N}$) leads to the \marked{asymptotic}
expression \marked{when $\ka\ll 1$} for the
contact potential (see Eq.~(D6) from~\cite{TT07})
\begin{equation}
  \label{eq:phi0}
  \phi_0=z_{+}^{-1}
  \ln\left[
    \frac{(\kappa a)^2 z_{+}}{2(z_{+}+z_{-})}
    \left(
    \frac{\sin(\mutilde\ln(\kappa a)+\Psi(\mutilde))}{\mutilde}
    \right)^2
    \right]
  \,,
\end{equation}
where $\mutilde$ is defined by $i\mutilde = 1+z_{+}A$ and
\begin{equation}
  \label{eq:Psi}
  \Psi(\mutilde)=-\frac{1}{2i}\ln\frac{z_{+} B^{-z_{+}}}{8(z_{+}+z_{-})(i\mutilde)^2}
  \,.
\end{equation}
This function has the property that $\Psi(0)=0$. To satisfy the
boundary condition $\lim_{r\to\infty} r\phi'(r)=0$ required by
electroneutrality, the constant of integration $B$ is a function of
$A$ and therefore of $\mutilde$. The explicit form of $B$ and
$\Psi(\mutilde)$ is only known in the cases of valencies 1:1, 1:2 and
2:1 where the connection problem of the long and short distances of
the PB solution has been solved~\cite{CTW77, TW98} and those results
are recalled in appendix~\ref{app:psi}. However, for other valencies,
we shall show that we only need to know the derivative at 0, ${\cal
  C}=\Psi'(0)$, and the definite integral ${\cal I}=\int_{-i}^{0}
\Psi(u)\,du$ to obtain information on the grand potential.

\begin{figure}
  \begin{center}
    \includegraphics[width=5cm]{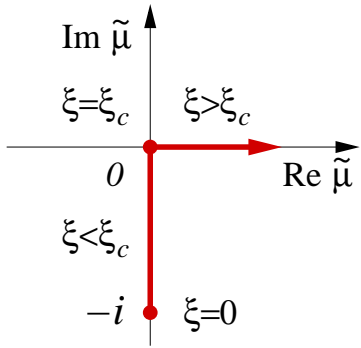}
    \caption{Path followed by the parameter $\mutilde$ on the complex plane $\mathbb{C}$ as $\xi$ increases. {For vanishing charge ($\xi=0$),
    $\widetilde \mu = -i$. Increasing $\xi$, $\widetilde \mu$ moves ``up'' along the imaginary axis, reaching $\widetilde \mu=0$ for
    $\xi=\xi_c$. For $\xi >\xi_c$, $\widetilde \mu$ is real and moves along the real axis; it is then a positive and non-decreasing function, see the arrow}.}
    \label{fig:mu}
  \end{center}
\end{figure}

The other constant of integration, $A$, and therefore $\mutilde$, are
obtained by applying the boundary condition at the contact of the
polyion $a\phi'(a)=2\xi$ which leads to
\begin{equation}
  \label{eq:xi-mu}
  z_{+}\xi-1 = \mutilde \cot\left(\mutilde\ln(\kappa a)+\Psi(\mutilde)\right)
  \,.
\end{equation}
For small charges $\xi\ll 1$ one has at order 0 {in} $\kappa a$,
$A=-\xi$ and $\mutilde=i(-1+z_{+}\xi)$. As $\xi$ increases, $\mutilde$ moves on the imaginary axis from $-i$ to $0$ where $\xi$ takes the critical value $\xi_c$ such that $\mutilde=0$, given by
\begin{equation}
  \label{eq:xic}
  z_{+} \xi_c-1= \frac{1}{\ln(\kappa a)+{\cal C}}
  \,.
\end{equation}
Notice that $\xi_c<1/z_{+}$: as explained in~\cite{TT06}, the effect
of salt is to reduce the condensation threshold from $1/z_{+}$ to
$\xi_c$.  For $\xi>\xi_c$, the parameter $\mu$ becomes real and moves
{along} the real axis up to the value~\cite{TT07}
\begin{equation}
  \label{eq:muinf}
  \mutilde_{\infty}=\frac{-\pi}{\ln(\ka)+{\cal C}}\,,
\end{equation}
as $\xi\to\infty$. The path followed by $\mutilde$ is shown in
Fig.~\ref{fig:mu}. In past works~\cite{TW97,TT06,TTexactPB06,TT07},
Eq.~(\ref{eq:phi0}) was exclusively used in the region $\xi\geq\xi_c$
($\mutilde\in\mathbb{R}$), however it should be clear from its
derivation~\cite{TW97, TTexactPB06, TT07} that Eq.~(\ref{eq:phi0}) is
also valid when $\xi<\xi_c$ provided that $\mutilde$ is imaginary.

%% Fig 1: mu path on complex plane

\marked{In this present work, we are interested in the situation with
  added salt. Nevertheless, it is interesting to comment on the
  similarities and differences of this case with the no-salt case,
  where there are only counterions in the solution. In
  Refs.~\cite{FKL51, ABM51}, the electrostatic potential in the
  no-salt case was worked out, which formally bears a close
  resemblance to Eq.~(\ref{eq:phi0}) with $a$ replaced by $r$ and when
  the integration constants are identified: in Eq.~(8) of
  Ref.~\cite{FKL51}, the integration constant $C$ plays the role of
  $\mutilde$ and the integration constant $\ln A$ (from
  Ref.~\cite{FKL51}) plays the role of $\Psi(\mutilde)$. The reason for
  this resemblance is that, at the close vicinity of the polyion, the
  coion density is negligible, thus the potential should behave
  asymptotically as the one of the no salt case.  However, at large
  distances from the polyion, the expression for the potential becomes
  more involved in the case with added salt, and Eq.~(\ref{eq:phi0}),
  with $a$ replaced by $r$, cannot be used at the edge of the
  Wigner-Seitz cell to determine the integration constant $\Psi(\mutilde)$
  by applying the electroneutrality boundary condition, contrarily to
  the development done in the no-salt case~\cite{FKL51}. Explicit
  expressions for this integration constant can only be obtained in
  the infinite dilution limit (infinite radius of the Wigner-Seitz
  cell) with the solution to the connection problem~\cite{CTW77,
    TW98}.  }

Several approximations have been developed for {solving}
Eq.~(\ref{eq:xi-mu})~\cite{TTexactPB06, TT06, TT07} depending on the
range of values of $\xi$ ($\xi<\xi_c$ or $\xi>\xi_c$). Therefore it
might prove difficult to perform the integral over $\xi$ in
Eq.~(\ref{eq:charging}), as pointed out in~\cite{S10}. 
{Attempting such a calculation is indeed inconvenient. However, a change of variable 
from $\xi$ to $\mutilde$ proves a useful reformulation. As we will show below, this leads to}
% It is indeed
% not wise to try to perform the integral over $\xi$ directly but rather to
% change to the variable $\mutilde$ where, as we will show below, 
an indefinite integral {that} can be computed independently of the range of values
considered for $\xi$.

First, we perform an integration by parts,
\begin{align}
  -\omega &= \int_0^{\xi} \phi_0(\xi')\,d\xi'=
  \left(\xi-\frac{1}{z_{-}}\right)\phi_0
  -\int_{0}^{\phi_0} \left(\xi-\frac{1}{z_{+}}\right) d\phi_0
  \nonumber\\
  &=\left(\xi-\frac{1}{z_{-}}\right)\phi_0
  -\int_{-i}^{\mutilde}
  \left(\xi-\frac{1}{z_{+}}\right)\frac{d\phi_0}{d\mu}\,d\mu
  \,,
  \label{eq:intphi1}
\end{align}
which can be interpreted as considering the thermodynamic potential
appropriate for a fixed potential polyion rather than a fixed charge
one~\cite{CM83}. In principle, the integral over $\mutilde$ should
follow the path shown in Fig.~\ref{fig:mu}, however this is
nonessential since the integrand is an holomorphic function of
$\mutilde$ in the vicinity of the path considered in the complex
plane. Now, from Eq.~(\ref{eq:phi0}), we have
\begin{equation}
  \frac{d\phi_0}{d\mutilde}
  =2 z^{-1}_{+} \left(
  (\ln(\ka)+\Psi'(\mutilde))
  \cot(\mutilde\ln(\kappa a)+\Psi(\mutilde))
  -\frac{1}{\mutilde}
  \right)
  \,.
\end{equation}
Using~(\ref{eq:xi-mu}), it is useful to notice that
\begin{equation}
  \frac{d}{d\mutilde}(z_{+}\xi-1)
  =
  \cot(\mutilde\ln(\ka)+\Psi(\mutilde))
  -\mutilde ( \ln(\ka)+\Psi'(\mutilde))\left(1+
  \cot^2(\mutilde\ln(\kappa a)+\Psi(\mutilde))\right)
  \,.
\end{equation}
With this, an exact differential appears in the integrand of (\ref{eq:intphi1})
\begin{equation}
  \left(\xi-\frac{1}{z_{+}}\right)\frac{d\phi_0}{d\mutilde}
  =
  -2z_{+}^{-1}
  \left[
    \frac{d}{d\mutilde}(z_{+}\xi -1)
    +\mutilde (\ln(\ka)+\Psi'(\mutilde))
    \right]\,.
\end{equation}
This yields
\begin{equation}
  \label{eq:omega1}
  \omega=
  -\left(\xi-\frac{1}{z_{+}}\right)\phi_0
  -\frac{2}{z_{+}}\xi
  -\frac{1}{z_{+}^2}
  \left[
    (\mutilde^2+1)\ln(\ka)
    +2\mutilde\Psi(\mutilde)+2i\Psi(-i)
    -2\int_{-i}^{\mutilde}\Psi(u)\,du
    \right]
  \,.
\end{equation}
The value of $\Psi(-i)$ corresponds to the situation of an uncharged polyion with $B=1$ in~(\ref{eq:Psi}). Then, $2i\Psi(-i)=-\ln(8z_{+}^{-1}(z_{+}+z_{-}))$. Because of relation~(\ref{eq:xi-mu}), the following identity
\begin{equation}
  (z_{+}\xi-1)^2 + \mutilde^2 =
  \left(
  \frac{\mutilde}{\sin(\mutilde\ln(\kappa a)+\Psi(\mutilde))}
  \right)^2
  \,,
  \label{eq:useful-eq}
\end{equation}
is satisfied. This can be used in (\ref{eq:phi0}) to obtain an alternative
expression of the contact potential $\phi_0$, that when replaced into (\ref{eq:omega1}) gives
\begin{align}
  \label{eq:omega2}
  \omega &=-
  \frac{1}{z_{+}}
  \xi
  \left(
    2+\ln\frac{(\ka)^2 z_{+}}{2(z_{+}+z_{-})}
    \right)
    +\frac{1}{z_{+}}\left(\xi-\frac{1}{z_{+}}\right)
    \ln[(z_{+}\xi-1)^2 + \mutilde^2]
    \nonumber\\
    &
    +\frac{1}{z_{+}^2}\left(
    2\ln 2+(1-\mutilde^2)\ln(\ka)-2\mutilde\Psi(\mutilde)
    +2\int_{-i}^{\mutilde} \Psi(u)\,du
    \right)
    \,.
\end{align}
This is the general exact analytic asymptotic expression for the grand
potential when $\kappa a \ll 1$ valid for all values of $\xi$. The
parameter $\mutilde$ is obtained by solving Eq.~(\ref{eq:xi-mu}). In
the following sections, we will develop some approximate solutions for
Eq.~(\ref{eq:xi-mu}) depending on the range of values of $\xi$ of
interest.

It is worth noticing that the grand potential is an holomorphic
function of $\mutilde$ in the vicinity of the path shown in
Fig.~\ref{fig:mu}, in particular close and at $\mutilde=0$ corresponding to
$\xi=\xi_c$. Therefore, in the strict sense, there isn't any phase
transition for any value of $\xi$ at any value of $\kappa a$. The
grand potential changes smoothly with $\xi$, even in the region close
to $\xi_c$ where the counterion condensation/de-condensation
occurs. However, the change from imaginary $\mutilde$ to real
$\mutilde$ does have quantitative implications on the small distance
behavior of the electrostatic potential as it has been analyzed
in~\cite{TTexactPB06}. For this reason, following tradition and
{with a slight abuse of language}, we will {refer to} $\xi_c$ as the ``critical''
value for counterion condensation.

\section{Results for moderate to highly charged polyions}
\label{sec:xi-close-xic}

In this section we develop a simplified expression of the previous
result~(\ref{eq:omega2}) that is valid for a wide range of charges
$\xi$ which includes all the region $\xi\geq\xi_c$ but also part of
the region below the critical value ($\xi<\xi_c$) provided $\xi_c-\xi
\ll 1$. This covers the most relevant range of values of $\xi$ for
physico-chemical and biological applications, including the
description of single (ss) and double stranded (ds) DNA where $\xi$
ranges between 2 and 4.2.

For this range of values of $\xi$, the parameter $\mutilde$ is
small. As in previous works, we perform a linearization of the
function $\Psi(\mutilde)\simeq {\cal C} \mutilde$. Then
Eq.~(\ref{eq:xi-mu}) can be written as
\begin{equation}
  \label{eq:xi-muhat}
  \frac{z_+ \xi-1}{z_+\xi_c-1}=\muhat \cot(\muhat)
  \,,
\end{equation}
where we defined
\begin{equation}
  \muhat=\mutilde (\ln(\ka)+{\cal C})= \frac{\mutilde}{z_+ \xi_c-1}
  \,.
\end{equation}
%% Inverse of mu cot(mu)
\begin{figure}
  \begin{center}
    \includegraphics[width=10cm]{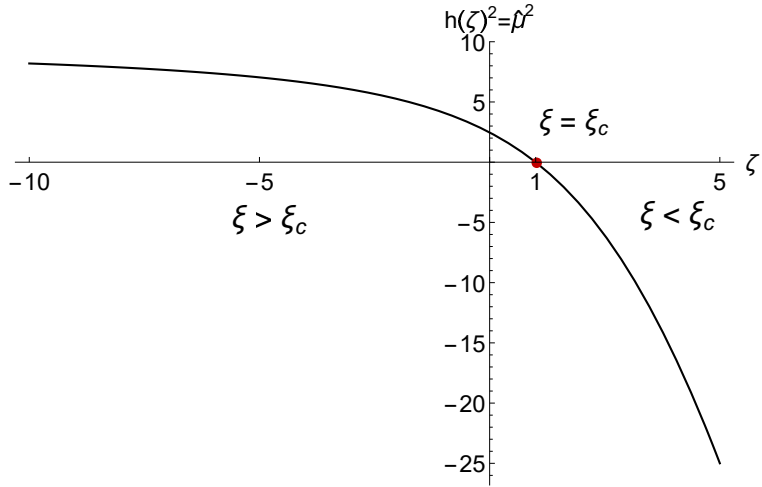}
  \end{center}
  \caption{The square of $h$, {defined as} 
  the inverse function of $z\mapsto z\cot(z)$, which gives the value of $\muhat^2$ as a function of $\zeta=(z_{+} \xi-1)/(z_{+}\xi_c-1)$.}
  \label{fig:h}
\end{figure}
Let $h$ be the inverse of the function $\muhat\mapsto
\muhat\cot(\muhat)$ which can easily be tabulated numerically
to any desired precision. Let
\begin{equation}
  \zeta=\frac{z_+ \xi-1}{z_+\xi_c-1}\,,
\end{equation}
then $\muhat=h(\zeta)$, so that $\mutilde$ is given by
\begin{equation}
  \label{eq:mutilde-sol}
  \mutilde=(z_+ \xi_c -1) h\left(\zeta\right)
  \,.
\end{equation}
It should be noted that $h$ has several branches because
Eq.~(\ref{eq:xi-muhat}) has an infinite number of solutions. We take
the branch such that $h(1)=0$ (corresponding to $\mutilde=0$ when
$\xi=\xi_c$). Also, $h$ is to be considered as a complex
function. When its argument $\zeta$ is real and $\zeta\leq 1$,
$h(\zeta)$ is real, ($\muhat^2=h(\zeta)^2\geq 0$, corresponding to $\xi
\geq\xi_c$ and $\mutilde\in[0,\mutilde_{\infty}])$. When $\zeta >1$,
$h(\zeta)$ takes imaginary values ($\muhat^2=h(\zeta)^2<0$,
corresponding to $\xi<\xi_c$ and
$\mutilde\in[-i,0]$). Fig.~\ref{fig:h} shows a plot of the square of
the function $h$ which gives the value of $\muhat^2$ as a function of
$\zeta$.

The linearization of the function $\Psi$ around $\mutilde=0$ leads to
the following approximation for the grand potential~(\ref{eq:omega2})
\begin{align}
  \label{eq:omega3}
  \omega &=-
  \frac{1}{z_{+}}
  \xi
  \left(
    2+\ln\frac{(\ka)^2 z_{+}}{2(z_{+}+z_{-})}
    \right)
    +\frac{1}{z_{+}}\left(\xi-\frac{1}{z_{+}}\right)
    \ln[(z_{+}\xi-1)^2 + \mutilde^2]
    \nonumber\\
    &
    +\frac{1}{z_{+}^2}\left(
    2\ln 2-\mutilde^2(\ln(\ka)+{\cal C})+\ln(\ka)
    +2{\cal I}
    \right)
    \,.
\end{align}
With $\mutilde$ obtained from Eq.~(\ref{eq:mutilde-sol}) (graphically
shown in Fig.~\ref{fig:h}), Eq.~(\ref{eq:omega3}) gives the grand
potential for $\xi$ in the range close to $\xi_c$ (both below and
above) and in all the range $\xi>\xi_c$ including highly charged
cylinders. If one is interested only in the {dependence} of the grand
potential on the salt concentration through the value of $\kappa a$,
it can be checked that Eq.~(\ref{eq:omega3}) reproduces the results
from Eq.~(13) of Ref.~\cite{S10} which gives the grand potential per
elementary charge ($g=\omega/\xi$) dependency on $\kappa a$ for a 1:1
electrolyte only. However, in that work, all the dependency of $g$ on
$\xi$ and on the electrolyte valencies was hidden in an arbitrary
reference value (named $G_{\text{ref}}^{\text{coul}}$ in
Ref.~\cite{S10}) which was inaccessible analytically up until now. Our
result, Eq.~(\ref{eq:omega3}), provides more complete results with the
complete $\xi$ dependence. It is also valid for other valencies
$z_{-}$:$z_{+}$ besides 1:1, provided two valency-dependent parameters
are known
\begin{eqnarray}
  {\cal C} &=& \Psi'(0),
  \label{eq:C}\\
  {\cal I} &=& \int_{-i}^0 \Psi(u)\,du\,.
  \label{eq:I}
\end{eqnarray}
In the cases 1:1, 1:2 and 2:1, these can be computed exactly,
\begin{eqnarray}
  {\cal C}_{1:1}&=&\gamma-3\ln 2 \simeq -1.50223 \\
  {\cal C}_{1:2}&=&\gamma-(3\ln 3)/2- (\ln 2)/3\simeq -1.30175\\
  {\cal C}_{2:1}&=&\gamma-(3\ln 3)/2- \ln 2\simeq -1.76385
\end{eqnarray}
and
\begin{eqnarray}
  {\cal I}_{1:1}&=& 1-6 \ln{\cal A} - (\ln 2)/3 \simeq -0.723576
  \label{eq:I11}\\
  {\cal I}_{1:2}&=& 1- 6 \ln{\cal A} - \ln 2 + (\ln 3)/2 \simeq -0.636368
  \label{eq:I12}\\
  {\cal I}_{2:1}&=& 1-6 \ln{\cal A} - \ln 2 + (5\ln 3)/16 \simeq -0.842358
  \label{eq:I21}\\
\end{eqnarray}
with $\gamma\simeq 0.577216$ the Euler Mascheroni constant and ${\cal
  A}\simeq 1.28243$ the Glaisher constant. The values of ${\cal C}$
were computed in~\cite{TTexactPB06} and the calculation of ${\cal I}$
is shown in appendix~\ref{app:psi}. For other valencies, we computed
numerically the values of ${\cal C}$ and ${\cal I}$ and the values are
reported in Table~\ref{tab:CyI}. Appendix~\ref{app:numerics} explains
the details of this numerical evaluation. {It should be kept in mind that up to a trivial 
rescaling, 1:1, 2:2, 3:3 etc. electrolytes are all equivalent within PB theory (only the ratio $z_+/z_-$ does matter).
Yet, upon increasing ionic valencies, correlation effects, discarded at the PB level,
become more prevalent and may invalidate the mean-field assumption \cite{Netz01}.}

\begin{table}
  \[\begin{array}{|c|c|c|c|}
\hline
 z_- & z_+ & {\cal C} & {\cal I} \\
\hline
 4 & 1 & -2.069 & -0.98620 \\
\hline
 3 & 1 & -1.938 & -0.92408 \\
\hline
 2 & 1 & -1.764 & -0.84236 \\
\hline
 3 & 2 & -1.649 & -0.78950 \\
\hline
1 & 1 & -1.502 & -0.72358 \\
\hline
 2 & 3 & -1.377 & -0.66861 \\
\hline
 1 & 2 & -1.302 & -0.63637 \\
\hline
 1 & 3 & -1.215 & -0.59988 \\
\hline
 1 & 4 & -1.167 & -0.57969 \\
\hline
\end{array}\]

  \caption{The constants ${\cal C}$ and ${\cal I}$ needed for the determination of the grand potential for different valencies.}
  \label{tab:CyI} 

\end{table}

\section{Results and benchmark of the analytic predictions}
\label{sec:results}

We benchmarked our analytic result against a {direct numerical computation of the free energy.
Details are given in Appendix \ref{app:numerics}. Two methods have been used for the numerical calculation:
either from Eq. \eqref{eq:charging} which requires to solve numerically a number of PB equations at fixed $\kappa a$ for
a number of charges (starting from $\xi=0$), or alternatively from Eq. \eqref{eq:checknew} which only requires
the solution of PB equation at the chosen values of $\xi$ and $\kappa a$.
Checking that both methods yield identical results is important for assessing the  validity of the calculations.}
% evaluation of the
% grand potential by solving numerically PB equation for a large range
% of values of $\xi$ then integrating numerically the surface
% potential. 
Details of the numerical resolution of PB equation are
given in appendix~\ref{app:numerics}. Fig.~\ref{fig:g11_and_err} shows
the grand potential per elementary charge, $g=\omega/\xi$, for the
case $\kappa a=0.1$ and valencies 1:1 using several approximations:
``analytic'' stands for Eq.~(\ref{eq:omega3}) with $\mutilde$ obtained
from Eq.~(\ref{eq:mutilde-sol}), DH is Debye-H\"uckel prediction
shown in Eq.~(\ref{eq:gDH}), TW is the prediction from
Ref.~\cite{TW97} recalled in appendix~\ref{app:xi_le_xic},
Eq.~(\ref{eq:omegaTW11}), and ``large $\xi$'' is Eq.~(\ref{eq:omega3})
with the approximation
\begin{equation}
  \label{eq:muapprox}
  \mutilde\simeq\frac{-\pi}{\ln(\ka)+{\cal C}+(z_{+}\xi-1)^{-1}} \,,
\end{equation}
which is valid asymptotically for large $\xi\gg\xi_c$~\cite{TT07}. The
inset shows the relative error between the different predictions
against the numerical calculation of the grand
potential. Interestingly, there is a large overlap between the simple
Debye-H\"uckel prediction
\begin{equation}
  \label{eq:gDH}
  g_{\text{DH}}=\xi \frac{K_0(\ka)}{\ka K_1(\ka)}
  \,,
\end{equation}
(where $K_0$ and $K_1$ are the modified Bessel functions of order 0
and 1), and the analytic prediction Eq.~(\ref{eq:omega3}) with
Eq.~(\ref{eq:mutilde-sol}). For large $\xi\gg\xi_c$,
Eq.~(\ref{eq:omega3}), with either Eq.~(\ref{eq:mutilde-sol}) or
Eq.~(\ref{eq:muapprox}) for the determination of $\mutilde$, provides
extremely accurate results with a relative difference between the
numerics and the analytic predictions that are below 0.2\%
($\xi_c\simeq 0.737175$ for $\kappa a = 0.1$). This is 10 times more
accurate than previous analytic predictions~\cite{S10} for that
range of values.

For $\xi<\xi_c$, the solution for $\mutilde$ {involves} the region
$\zeta>1$ of $h(\zeta)$ (see Figure~\ref{fig:h}). In this region
$|\muhat|$ increases faster, therefore the range of validity of the
approximation $|\mutilde|\ll 1$ is smaller. Nevertheless, the
analytic prediction of Eq.~(\ref{eq:omega3}) with
Eq.~(\ref{eq:mutilde-sol}) remains accurate for values of $\xi$
smaller than $\xi_c$ provided $\xi_c-\xi\ll 1$.  In practice,
Eq.~(\ref{eq:omega3}) can be applied down to values of $\xi=1/(2
z_{+})$ (half the Manning parameter $1/z_{+}$) with an error that
starts to become larger than $1\%$ below that threshold. Then, for
smaller values of $\xi$ the most accurate analytic expression is
provided by the Debye-H\"uckel prediction~(\ref{eq:gDH}). It turns out
that the analytic expression from Ref.~\cite{TW97} recalled in
Eqs.~(\ref{eq:omegaTW11})-(\ref{eq:omegaTW21}) is less accurate than
the DH prediction. This is probably traced back to the fact that DH
prediction gives the correct asymptotics for $\xi\to 0$ of the
nonlinear PB problem regardless of the value of $\kappa a$ (it is not
limited to $\ka \ll 1$). In summary, our main result
Eq.~(\ref{eq:omega3}) with Eq.~(\ref{eq:mutilde-sol}) for moderate to
highly charged polyion, combined with DH prediction~(\ref{eq:gDH}) for
smaller values of $\xi$ provide excellent accurate analytic
predictions for the grand potential.

%%% Figure Error: analytic 11 vs numeric
\begin{figure}
  \begin{center}
    \includegraphics[width=10cm]{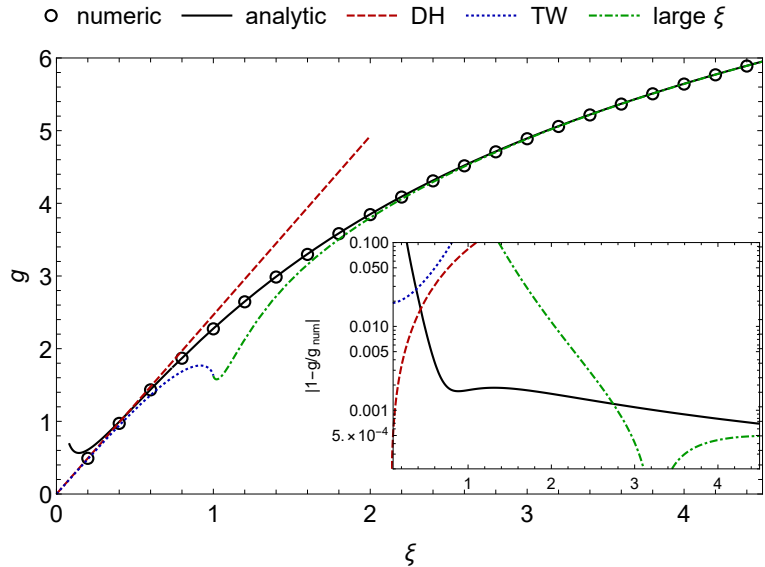}
  \end{center}
  \caption{Comparison of the different predictions for the grand
    potential $g$ for $\kappa a=0.1$ and 1:1 electrolyte as a function of
    $\xi$: ``analytic'' is Eq.~(\ref{eq:omega3}) with
    Eq.~(\ref{eq:mutilde-sol}), ``DH'' is Debye-H\"uckel prediction
    Eq.~(\ref{eq:gDH}), ``TW'' is Tracy and Widom prediction
    Eq.~(\ref{eq:omegaTW11}) and ``large $\xi$'' is
    Eq.~(\ref{eq:omega3}) but with $\mutilde$ approximated as shown in
    Eq.~(\ref{eq:mutilde-sol}). Inset: Relative error between the
    analytic predictions and the numeric result of the grand
    potential.}
  \label{fig:g11_and_err}  
\end{figure}

\begin{figure}
  \begin{center}
    \includegraphics[width=8cm]{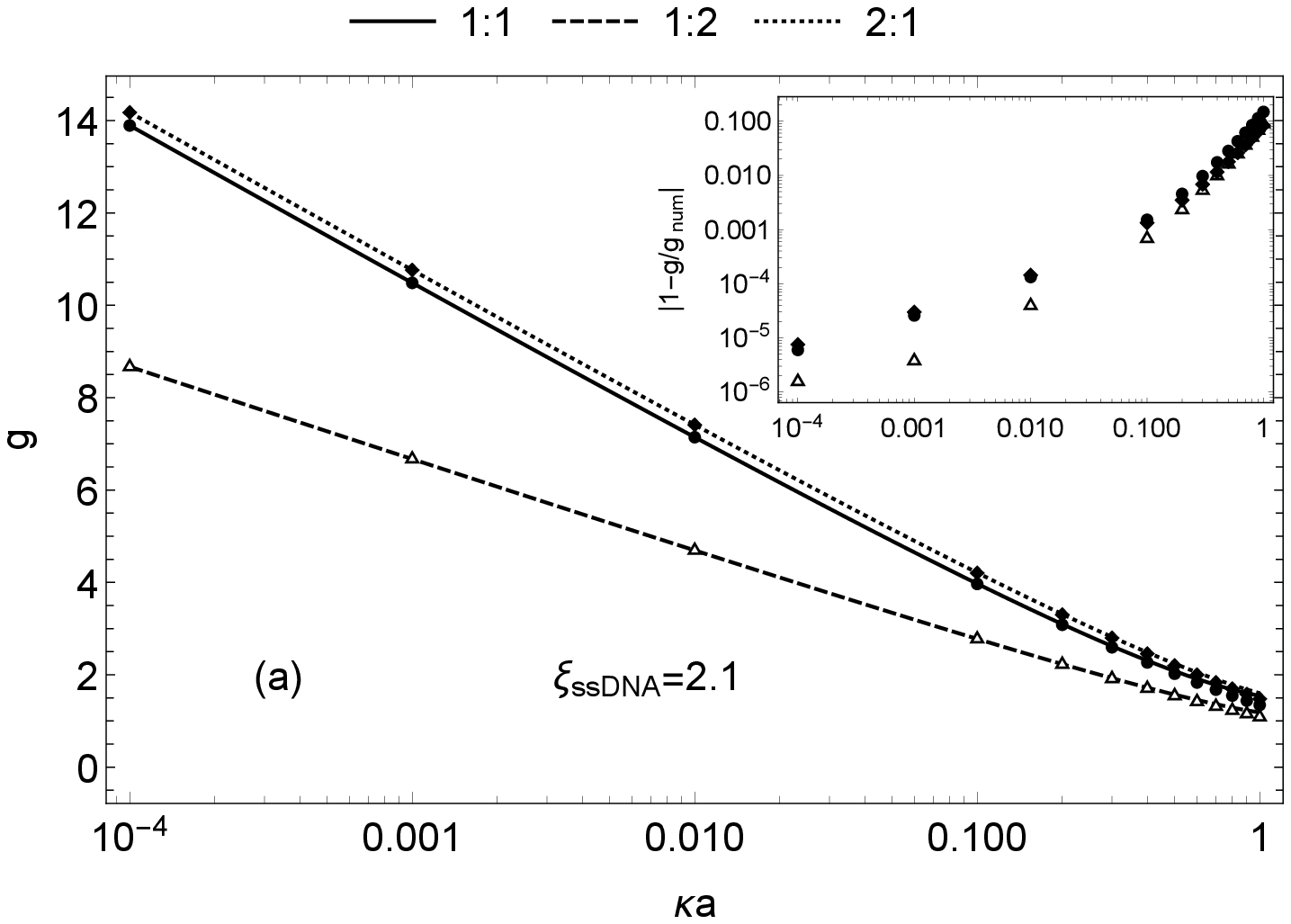}
    \includegraphics[width=8cm]{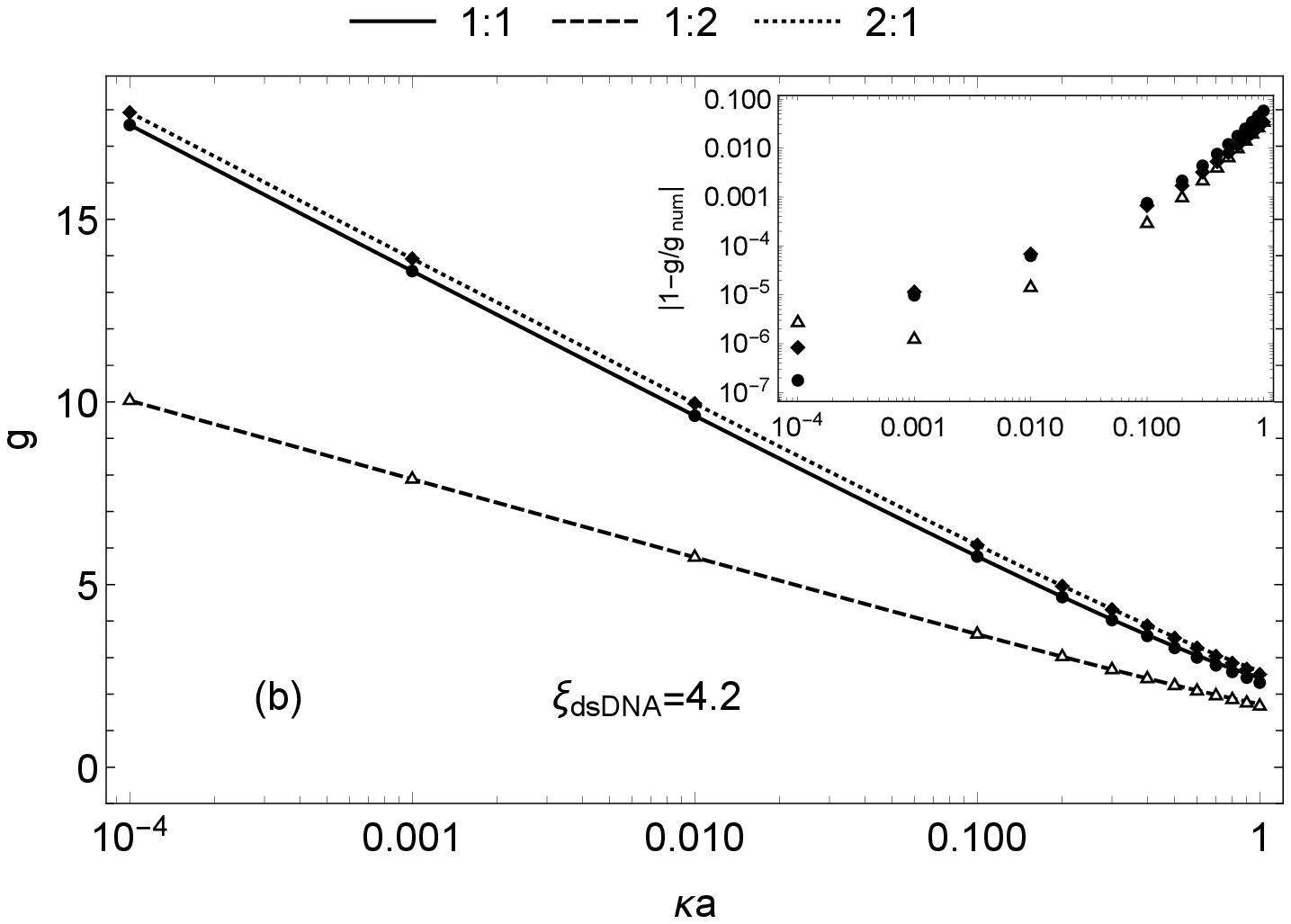}
  \end{center}
  \caption{Grand potential as a function of $\kappa a$ for (a) ssDNA and (b) dsDNA. The lines are the analytic predictions and the symbols the numerical evaluation. Inset: relative error between the analytic prediction and the numerical evaluation. {The grand potential is computed per elementary charge $g=\omega/\xi$.} }
  \label{fig:gDNA}
\end{figure}

To test the accuracy of our prediction when $\kappa a$ varies, we
consider two experimentally relevant cases corresponding to single
stranded DNA {$\xi=\xi_{\text{ssDNA}}=2.1$} and double stranded DNA
{$\xi=\xi_{\text{dsDNA}}=4.2$}. Fig.~\ref{fig:gDNA} shows the grand
potential as a function of $\kappa a$ (obtained varying salt
concentration) with a comparison to the numerical evaluation. The
insets of the figures show the relative error between the two. The
worst case is for 1:1 electrolyte and ssDNA where the error reaches
values beyond 1\% {but} only for $\kappa a\geq 0.4$. 
% for ssDNA and 1:1 electrolyte. 
Even when $\kappa a = 1$, where the analytic treatment
is not supposed to be {accurate}, we obtained a fair approximation for the
grand potential with relative error below 3.5\% for a 1:2 and 2:1
electrolytes for dsDNA.

\begin{figure}
  \begin{center}
    \includegraphics[width=10cm]{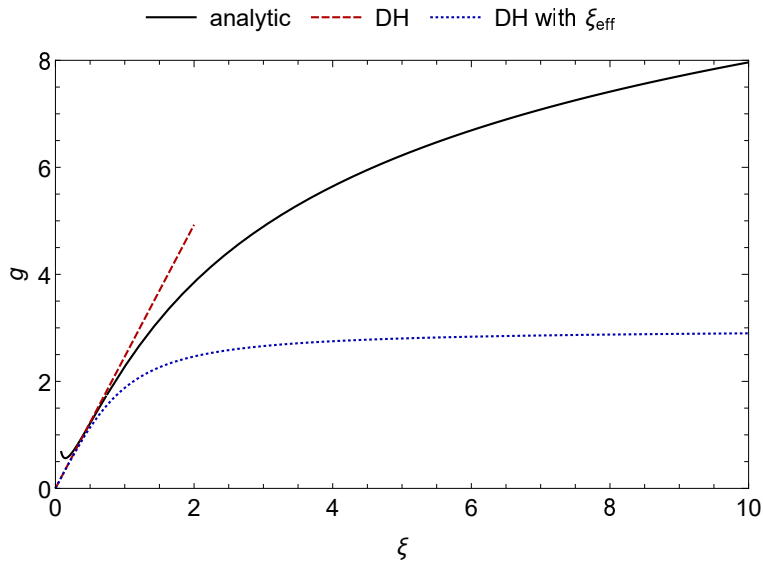}
  \end{center}
  \caption{Failure of the prediction for the grand potential $g$ using
    the Debye-H\"uckel theory combined with effective charge
    corrections $\xi_{\text{eff}}$. The data is shown for $\kappa a
    =10^{-1}$ and 1:1 electrolyte and compared to the analytic
    results Eq.~(\ref{eq:omega3}).}
  \label{fig:DH_eff_fail}
\end{figure}
For highly charged polyions, the Debye-H\"uckel theory has often been
applied by correcting the bare charge with the effective one which
encodes the large distance features of the electric
potential~\cite{ACGMPH84, TBAG03, ATB03,
  TTexactPB06}. Fig.~\ref{fig:DH_eff_fail} compares the analytic
result with this prescription, showing that, for the free energy and
grand potential calculations, the effective charge concept {as previously formulated} fails. In
particular, since the effective charge saturates for highly charged
polyions, it would predict that the grand potential per elementary
charge will saturate, when in reality this is not the case. Our
analytic results~(\ref{eq:omega3}) predict that for highly polyions,
the grand potential behaves as
\begin{equation}
  \label{eq:g-large-xi}
  g=\frac{1}{z_{+}} \left( 2\ln \xi
  -2-\ln\frac{(\ka)^2}{2(z_+ + z_-)z_+}\right) +O(\xi^{-1}\ln\xi)
  \,,\qquad \xi\to\infty
  \,.
\end{equation}
The failure of the effective charge prescription for the computation
of the grand potential is probably due to the strong free energy
contribution of the condensed ions {i.e.~of short scale features ignored by the far-field behaviour 
subsumed in the  effective charge.}
% that the far field potential fails to keep trace.

\begin{figure}
  \begin{center}
    \includegraphics[width=10cm]{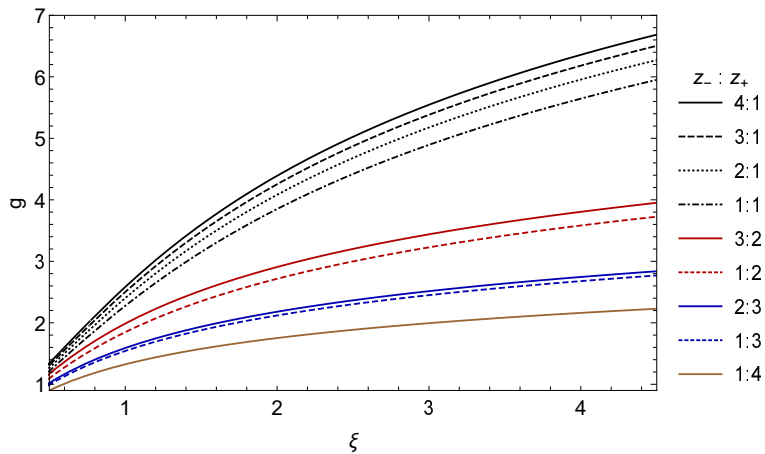}
  \end{center}
  \caption{Grand potential per elementary charge $g$ as a function of
    $\xi$ for different valencies $z_-$:$z_+$ at $\kappa a=10^{-1}$. }
  \label{fig:g_zm_zp_vs_xi_ka0_1}
\end{figure}

\begin{figure}
  \begin{center}
    \includegraphics[width=10cm]{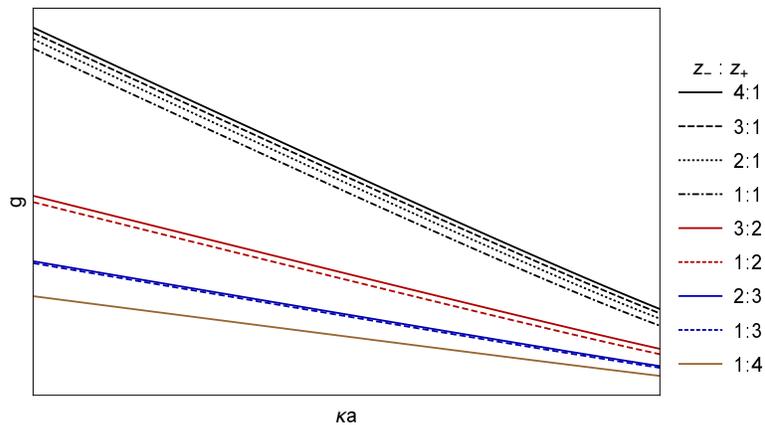}
  \end{center}
  \caption{Grand potential per elementary charge $g$ as a function of
    $\kappa a$ for different valencies $z_-$:$z_+$ at
    {$\xi=\xi_{\text{dsDNA}}=4.2$}. }
  \label{fig:g_zm_zp_vs_ka_dsDNA}
\end{figure}

We now discuss the dependency of the grand potential on the
valencies $z_{-}$:$z_{+}$ of the electrolyte. Using our prediction
Eq.~(\ref{eq:omega3}) and the data from Table~\ref{tab:CyI} obtained
in appendix~\ref{app:numerics}, we plot $g$ as a function of $\xi$
when the valency is changed (Fig.~\ref{fig:g_zm_zp_vs_xi_ka0_1}). The
salt concentration is fixed {at} $\kappa a=10^{-1}$. Note how the
different curves can be regrouped by common counterion valency
$z_{+}$. At fixed counterion valency ($z_{+}$), when the coion valency
$z_{-}$ is increased, the grand potential increases moderately. On the
other hand, an increase on the counterion valency $z_{+}$ reflects
 in a large decrease on the grand potential. This is also
{apparent} on the analytic expression~(\ref{eq:omega3}), where it can
be appreciated that the dependency on $z_{-}$ is logarithmic while
there are terms proportional to $z_{+}$ and $z_{+}^2$ responsible of a
stronger dependency on $z_{+}$ than on
$z_{-}$. Fig.~\ref{fig:g_zm_zp_vs_ka_dsDNA} confirms this trend,
showing now $g$ for dsDNA as a function of $\ka$ for different
valencies. From Eqs.~(\ref{eq:omega3}) and (\ref{eq:g-large-xi}), one
can notice that at fixed large $\xi$, the leading behavior dependence
on $\ka$ is $g\sim -(2/z_{+}) \ln (\ka)+ O(1)$ for $\ka \ll 1$. This
linear dependence of $g$ on $\ln (\ka)$ at leading order is verified
on Fig.~\ref{fig:g_zm_zp_vs_ka_dsDNA}, where the slope of the curves
in log-scale is indeed $-2/z_{+}$. The coion valency $z_{-}$ only
affects the subleading order terms.

\section{Summary and conclusion}

{Coulomb interactions are key to rationalizing the thermodynamics 
of nucleic acid processes \cite{S10}, or other properties of biopolymers such as
their persistence length \cite{ShTr16}. The non-linear Poisson-Boltzmann (PB) theory adopted here 
is a mean-field framework that provides a useful description, not only in the present 
context but more generally for studying soft matter in aqueous solutions, where Coulombic
effects are paramount \cite{Levin02,Andelman06,Messina09}.}
We computed the exact analytic low-salt asymptotic expansion of the
grand potential{/free energy} of a {long} cylindrical polyion 
(Eqs.~(\ref{eq:omega3}) and (\ref{eq:mutilde-sol})). {The biopolymer is thus modeled here 
are as a uniformly charged straight cylinder, and we addressed the case of a binary electrolyte,
with arbitrary co/counter ion valencies  $z_-$ and $z_+$.
Analytical progress} was possible
{taking advantage} of the contact potential (\ref{eq:phi0}) derived in previous
works~\cite{TT07} and {writing expressions valid} for all values of {polyion charge} $\xi$.
{This required to introduce an auxiliary quantity, $\mutilde$,  appropriately allowed to take complex values,
either pure imaginary or real, as sketched in} Fig.~\ref{fig:mu}. This {results in a significantly extended range} of
validity of the contact surface potential and the quantities derived
from it, such as the preferential interaction coefficient computed in
Ref.~\cite{TT07}, and the grand potential computed here. With this, one
can obtain reliable results for moderate to highly charged polyions,
{having} linear charge $\xi$ {larger than} half the Manning
parameter {($\xi>1/(2 z_+)$, therefore $1/2$ for monovalent counterions)}. {The regime of smaller charges is somewhat less interesting:}
for smaller values {of $\xi$}, one enters the {realm} of the simple
linear Debye--H\"uckel theory, {which provides} accurate results. This opens
the opportunity to {present} the analytic results for the cylindrical PB
equation~\cite{TW97, TT06, TTexactPB06, TT07} in a unified framework,
{that no longer requires to consider} different formulas for $\xi<\xi_c$ and
$\xi\geq\xi_c$, {as done in previous publications. In practice, our low salt approach turns reliable 
for $\kappa a <1$. We finally emphasize that our work sheds some light} 
into the analytic properties 
%({in the mathematical sense}) 
of the grand potential. {Within the present PB formalism, it} turns out to be a holomorphic
function of $\mutilde$ and $\xi$, {with no singularity} even at $\xi=\xi_c$,
{unlike what the widespread terminology pertaining to Manning ``condensation transition'' may lead to believe}.

\begin{acknowledgments}
  This work is partially funded by ECOS-Nord action
  C18P01. G.T. acknowledges support from Fondo de Investigaciones,
  Facultad de Ciencias, Universidad de los Andes, Research Program
  2018-2019 ``Modelos de baja dimensionalidad de sistemas cargados''.
\end{acknowledgments}

\appendix

\section{Numerical evaluation of the grand potential}
\label{app:numerics}

In this appendix we give a few details {for} algorithm to compute the grand potential{/free energy}. The numerical
resolution of PB equation was done with {\sc Mathematica} based on the
code presented in Appendix A of Ref.~\cite{TBAG03}. Essentially,
PB equation is solved on a cylindrical Wigner-Seitz cell of large
radius $R=22 \kappa^{-1}$ with boundary conditions at the edge of the
cell $\phi'(R)=0$ (by electroneutrality) and a test value for the
potential at the edge $\phi(R)=\phi_{\text{edge}}$. The resolution of
the differential equation is done with {\sc Mathematica}
\texttt{NDSolve} built-in algorithm with options
\texttt{MaxSteps}$\to500$ and \texttt{WorkingPrecision}$\to33$.
{We are interested in the $R\to\infty$ limit (infinite dilution)
and $R$ should be chosen accordingly, large enough to provide an acceptable solution.
It is then necessary to check that the results obtained do not depend on $R$,
within the targeted accuracy}.

If the algorithm converges successfully, the corresponding linear
charge can be obtained as $\xi=a\phi'(a)/2$. In the linear regime, for
$\xi\ll \xi_c$, $\xi$ changes linearly with $\phi_{\text{edge}}$,
whereas in the nonlinear regime, for $\xi>\xi_c$, a small change on
$\phi_{\text{edge}}$ produces exponentially large changes on $\xi$.
Due to the potential saturation effect~\cite{TT03}, if
$\phi_{\text{edge}}$ is too large (beyond its saturation value
$\phi_{\text{sat}}$), the algorithm will not converge. By trial and
error, the saturation value of $\phi_{\text{edge}}$ can be
determined. A sweep over values of $\phi_{\text{edge}}$ provides a
table of data for the contact potential $\phi(a)$ and the
corresponding linear charge value $\xi$. This sweep should be done
with equal spacing \textit{on a log scale} of values of
$\phi_{\text{edge}}$ starting at $\phi_{\text{sat}}$ to account for
the saturation effect to obtain linearly evenly spaced for values of
$\xi$. In practice, we used $\phi_{\text{edge}}=\phi_{\text{sat}} (1 -
(97/100)^k)$ with $k$ ranging from 0 to 400 by step increments of
5. This produces a table of 80 values of the contact potential and its
corresponding linear charge. This table is interpolated to produce
numerically the function $\xi\mapsto\phi_0$. The interpolation was
made with {\sc Mathematica} \texttt{Interpolation} function with
default options (degree 3 polynomial interpolation between successive
data points). This function is then integrated numerically with {\sc
  Mathematica} \texttt{Integrate} to obtain the grand potential
(Eq.~(\ref{eq:charging})). The above procedure is {followed} for a given
value of $\kappa a$ and gives the grand potential for any value of
$\xi$. If $\kappa a$ is changed the procedure should be run again
since the saturation value $\phi_{\text{sat}}$ changes.
{Besides, an important test for the correctness of the calculation is to check that the grand potential
values are recovered by a direct calculation, without any integral over $\xi$, making use 
of relation \eqref{eq:checknew} below. Our results satisfied this test}.

The numerical evaluation of the constants ${\cal C}$
(Eq.~(\ref{eq:C})) and ${\cal I}$ (Eq.~(\ref{eq:I})) for different
valencies $z_{-}$:$z_{+}$ was done as
follows. Using Eq.~(\ref{eq:useful-eq}) into Eq.~(\ref{eq:phi0}), shows that
the contact potential $\phi_0$ and the parameter $\mutilde$ satisfy
\begin{equation}
  e^{-z_{+}\phi_0} \frac{(\ka)^2 z_{+}}{2(z_{+}+z_{-})} -
  (z_{+}\xi-1)^2=\mutilde^2
  \,.
  \label{eq:zerofnct}
\end{equation}
Therefore, the left-hand side (LHS) of Eq.~(\ref{eq:zerofnct})
vanishes when $\xi=\xi_c$ ($\mutilde=0$). Along with the numerical
computation of the grand potential explained above, a data table of
the LHS of Eq.~(\ref{eq:zerofnct}) as a function of $\xi$ can be build
then interpolated. The zero of this interpolated function the closest
to $1/z_{+}$ is then found using {\sc Mathematica} \texttt{FindRoot}
algorithm to obtain $\xi_c$. With $\xi_c$ determined numerically, the
constant ${\cal C}$ is obtained from Eq.~(\ref{eq:xic}). The constant
${\cal I}$ is obtained from Eq.~(\ref{eq:omega3}) evaluated at
$\xi=\xi_c$ (corresponding to $\mutilde=0$). A strong test of this
algorithm is that it should give the same values of ${\cal C}$ and
${\cal I}$ independently of the chosen value of $\kappa a$ provided
it is small enough. We tested this numerical procedure using values of
$\kappa a=10^{-6}$, $10^{-5}$ and $10^{-4}$, confirmed the stability
of the numerical values of ${\cal C}$ and ${\cal I}$ and reproduced
the analytically known values for the solvable cases of valency 1:1,
1:2 and 2:1 with a relative accuracy of $10^{-5}$ for ${\cal C}$ and $10^{-6}$
for ${\cal I}$. Table~\ref{tab:CyI} provides the values of ${\cal C}$
and ${\cal I}$ for other valencies of experimental interest, bearing
in mind the limitations of PB framework for larger valencies \cite{Netz01}.

\section{Charging process to obtain the free energy}
\label{app:free-energy}

Several charging {processes} can be put forward to compute the free
energy by studying its variations with respect to different
parameters. A review of such process can be found in \marked{\cite{M55, SD89,
  TH96, TH97}}. Consider the cell model and PB theory for a
polyion. In~\cite{TH96} it is shown that the variations of the free
energy $F$ are given by
\begin{equation}
  \delta(\beta F) =
  \frac{1}{8\pi l_B} \oint_{\Sigma} (\phi \nabla(\delta \phi)
  - \delta \phi \nabla \phi)\cdot dS
  + \beta U \frac{\delta l_B}{l_B}
  +\int_{\cal P} \phi\, \delta\left(\frac{\sigma}{e}\right)\, dS
  + \sum_{s=\pm}\ln(n^0_{s}\Lambda_{s}^3)\delta N_{s}
  \,,
\end{equation}
where  ${\cal P}$ is
the surface of the polyion, $U$ the internal energy, $N_{\pm}$ the
number of positive and negative ions, $\sigma$ the surface charge
density of the polyion, and {$\Sigma$ is the surface of the Wigner-Seitz cell of arbitrary shape that encloses the system; in the present situation
of infinite dilution,
$\Sigma$ is ``sent to infinity'' and the corresponding integral is absent from the equation,
as a consequence of screening ; the spatial integrals considered consequently run over all space.} 
% In the present situation, the surface term on the cell vanishes because of the boundary conditions, and 
{Besides, the bulk electrolyte plays the role of a reservoir, with given chemical potentials
$\mu_\pm$ for cations and anions. It is thus} 
appropriate to work in the grand canonical ensemble due to the
chemical equilibrium with the salt reservoir. Therefore, we consider
the grand potential $\Omega=F- \mu_{+}N_{+} - \mu_{-}N_{-}$ and its
excess value with respect to that of the reservoir
$\Omega_0=\int(n_0^{+} + n_{0}^{-}) dV$.  At fixed $l_B$, its
variations are
\begin{equation}
  \label{eq:omega-vars-gen}
  \delta(\beta (\Omega-\Omega_0))=
  \int_{\cal P} \phi \,\delta\left(\frac{\sigma}{e}\right)\, dS -
  \int (n^{+}(\r)-n^{+}_0) \delta\mu_{+} dV
  +  \int (n^{-}(\r)-n^{-}_0) \delta\mu_{-} dV
  \,,
\end{equation}
where $n^{\pm}(\r)$ is the ionic density profiles around the polyion.
The second and third term show that the variations with respect to the
chemical potentials are the excess ionic charge around the polyion,
which is essentially the preferential interaction
coefficient~\cite{STR02, TT07}. Therefore, the grand potential can be
obtained by integrating the preferential interaction coefficient with
respect to the chemical potential
$\mu_{\pm}=\ln(n_{0}^{\pm}\Lambda^3_{\pm})$ or equivalently with
respect to $\ln(\ka)$. This strategy was used in~\cite{S10} to obtain
analytic predictions for the grand potential. However, it has the
disadvantage that it requires the determination of an arbitrary
reference value of the grand potential at a given salt density (0.15 M
was used in~\cite{S10}). This reference value is different for each
value of $\xi$ even if $\kappa a$ is kept fixed. In this work, we
followed {another} route by considering variations of the surface charge
$\sigma$ of the polyion. This is equivalent to {varying} $\xi$ since
$\sigma=-\xi e/(2\pi a l_B)$. For a cylinder of length $L$, at fixed
chemical potentials, Eq.~(\ref{eq:omega-vars-gen}) becomes
\begin{equation}
  \label{eq:omega-vars}
  \delta (\beta (\Omega-\Omega_0))
  =
  -\frac{L}{l_B} \phi_0 \,\delta \xi
  \,.
\end{equation}
Let us define the dimensionless excess grand potential per unit length
\begin{equation}
  \label{eq:def-omega}
  \omega = \beta (\Omega-\Omega_0) l_B/L\,,
\end{equation}
and the dimensionless excess grand potential per elementary charge
\begin{equation}
  \label{eq:def-g}
  g=\beta {(\Omega -\Omega_0)}/N = \omega/\xi
  \,,
\end{equation}
with $N=L/b$ the number of elementary charges of the polyion. Since at
$\xi=0$ the grand potential is $\Omega_0$, 
we obtain~(\ref{eq:charging}) from~(\ref{eq:omega-vars}).

{For putting to the test the reliability of our numerical solution, we have computed the
free energy/grand potential through an alternative route, that does not require any $\xi$-integration.
Once $\xi$ and $\kappa a$ have been chosen and PB equation solved, we have \cite{TH96}
\begin{equation}
  \beta \omega  \,=\, -\frac{1}{2} \,\xi\, \phi_0 
  +\,\frac{\kappa^2}{8 \pi (z_++z_-)} \left(
  \int \phi \left( e^{z_- \phi} - e^{-z_+ \phi} 
 \right)\, d^2V -2 \int \left[ \frac{1}{z_-}(e^{z_- \phi}-1) + \frac{1}{z_+} (e^{-z_+ \phi}-1)
 \right] \, d^2V \right),
 \label{eq:checknew}
\end{equation}
where the second integral on the right hand side is a rewriting of $\int (n^++n^- - n_0^+-n_0^-) d^2V$.
The two integrals in \eqref{eq:checknew} run over the 2D plane perpendicular to the cylinder axis, outside the charged cylinder
($r>a$).
}

\section{Explicit solutions for 1:1, 1:2 and 2:1 valencies}
\label{app:psi}

In the short distance asymptotics of the electrostatic potential from
which the contact potential is deduced (Eq.~(\ref{eq:phi0})),
$\mutilde$ and $\Psi(\mutilde)$ are the two constants of integration
of the differential equation~(\ref{eq:PB}). However, to satisfy the
boundary condition $r\phi'(r)\to 0$ when $r\to\infty$, the so-called
connection problem between the short and large scale behavior of the
solution has to be solved to find the relationship between the two
integration constants. This problem was {worked out} in the integrable cases
of valencies 1:1, 1:2, 2:1, where the solution to PB equation can be
written down in terms of Fredholm determinants~\cite{CTW77,TW98}. We
recall here {the main} results. For those valencies, the constant $B$
appearing in Eq.~(\ref{eq:Psi}) is given by~\footnote{In
  Ref.~\cite{TT07} the cases 1:2 and 2:1 were inverted in Eq. (D2).}
\begin{eqnarray}
  \label{eq:B11}
  B_{1:1}&=&2^{6A}\gamma\left(\frac{1+A}{2}\right)^2 \\
  \label{eq:B12}
  B_{1:2}&=&3^{3A}2^{2A}\gamma\left(\frac{1+2A}{3}\right)
  \gamma\left(\frac{2+A}{3}\right)\\
  \label{eq:B21}
  B_{2:1}&=&3^{3A}2^{2A}\gamma\left(\frac{2(1+A)}{3}\right)
  \gamma\left(\frac{1+A}{3}\right)
\end{eqnarray}
where $\gamma(z)=\Gamma(z)/\Gamma(1-z)$ with $\Gamma$ the Euler gamma
function. With $A$ related to $\mutilde$ by $i\mutilde=1+z_{+}A$,
replacing this in Eq.~(\ref{eq:Psi}) gives the function $\Psi$ 
\begin{eqnarray}
  \Psi_{1:1}(\mutilde)&=&-3\mutilde\ln 2
  + i \ln\left(\frac{i\mutilde}{2}
  \gamma\left(\frac{i\mutilde}{2} \right)\right)
  \\
  \Psi_{1:2}(\mutilde)&=&-\frac{\mutilde}{2}(3\ln3 +2\ln 2)
  +i\ln\left(\frac{i\mutilde}{3}\gamma\left(\frac{i\mutilde}{3}\right)\right)
  +i\ln\gamma\left(\frac{i\mutilde+3}{6}\right)
  \\
  \Psi_{2:1}(\mutilde)&=&-\frac{\mutilde}{2}(3\ln3 +2\ln 2)
  +\frac{i}{2} \ln\left(\frac{i\mutilde}{3}\gamma\left(\frac{i\mutilde}{3}\right)\right)
  +\frac{i}{2} \ln\left(\frac{2i\mutilde}{3}\gamma\left(\frac{2i\mutilde}{3}\right)\right)
  \,.
\end{eqnarray}
With this, the constant ${\cal I}=\int_{-i}^0
\Psi(\mutilde)\,d\mutilde$ used in Eq.~(\ref{eq:omega3}) can be
computed explicitly, {leading to} Eqs.~(\ref{eq:I11})-(\ref{eq:I21}).

\section{Results for $\xi$ below the critical value $\xi_c$}
\label{app:xi_le_xic}

Section~\ref{sec:xi-close-xic} main result Eq.~(\ref{eq:omega3})
breaks down if $\xi\ll\xi_c$. We develop here an approximation
appropriate for that range. When $\xi<\xi_c$, $\mutilde\in[-i,0]$,
therefore it is useful to introduce $\nu\in[-1,0]$ defined by
$\mutilde=i\nu$. Eq.~(\ref{eq:xi-mu}) becomes
\begin{equation}
  \label{eq:xi-nu}
  z_{+}\xi -1 = \nu \coth(\nu \ln(\ka)+\varphi(\nu))\,,
\end{equation}
with $\varphi(\nu)=-i\Psi(\mutilde)\in\mathbb{R}$. For $\kappa a\ll
1$, the argument of the hyperbolic cotangent in (\ref{eq:xi-nu}) is
large and positive. Therefore,
\begin{equation}
  z_{+}\xi-1=\nu (1+ 2e^{-2\nu\ln(\ka)+\varphi(\nu)} + o((\ka)^{-2\nu})
  \,,
\end{equation}
which yields
\begin{equation}
  \nu=(z_{+}\xi-1) (1-2(\ka)^{2(1-z_{+}\xi)} e^{-2\varphi(z_{+}\xi-1)}
  +o((\ka)^{2(1-z_{+}\xi)}) \,.
\end{equation}
Replacing this in Eq.~(\ref{eq:omega2}) and using Eq.~(\ref{eq:Psi}) we find
\begin{equation}
  \omega = -\xi^2 \ln(\ka) - \frac{1}{z_{+}} \int_{-1}^{-1+z_{+}\xi} \ln B\,
  d\nu
  \,,
\end{equation}
where $B$ should be seen as a function of $\nu=-(1+z_{+} A)$. This is
the same result as if one starts with the approximation $\phi_0=-2 A
\ln(\ka)+\ln B$, with $A$ approximated as $A=-\xi$. Further progress
can be only made in the cases of valencies 1:1, 1:2 and 2:1, where $B$
is explicitly known~\cite{CTW77, TW98}~(see
Eqs.~(\ref{eq:B11})-(\ref{eq:B21}) from appendix~\ref{app:psi}),
recovering previous results from {Tracy and Widom} Ref.~\cite{TW97},
\begin{eqnarray}
  \label{eq:omegaTW11}
  \omega_{1:1}
  &=&
      \xi^2(-\ln(\kappa a)+3\ln 2)
      +4[\psi^{(-2)}({\scriptstyle\frac{1-\xi}{2}})
      +\psi^{(-2)}({\scriptstyle\frac{1+\xi}{2}})
      -2\psi^{(-2)}({\scriptstyle\frac{1}{2}})]
      \,,\\
  \label{eq:omegaTW12}
  \omega_{1:2}
  &=&
      \xi^2(-\ln(\kappa a)+\frac{3}{2}\ln 3+\ln 2)
      +3\left[\frac{1}{2}(\psi^{(-2)}({\scriptstyle\frac{2(1+\xi)}{3}})
      +\psi^{(-2)}({\scriptstyle\frac{1-2\xi}{3}}))
      +\psi^{(-2)}({\scriptstyle\frac{1+\xi}{3}})
      +\psi^{(-2)}({\scriptstyle\frac{2-\xi}{3}})
      \right.
  \nonumber\\
  && \left.
     -\frac{3}{2}(
      \psi^{(-2)}({\scriptstyle\frac{2}{3}})+
      \psi^{(-2)}({\scriptstyle\frac{1}{3}}))\right]
      \\
      \label{eq:omegaTW21}
  \omega_{2:1}
  &=&
      \xi^2(-\ln(\kappa a)+\frac{3}{2}\ln 3+\ln 2)
      +3\left[\frac{1}{2}(\psi^{(-2)}({\scriptstyle\frac{2(1-\xi)}{3}})
      +\psi^{(-2)}({\scriptstyle\frac{1+2\xi}{3}}))
      +\psi^{(-2)}({\scriptstyle\frac{1-\xi}{3}})
      +\psi^{(-2)}({\scriptstyle\frac{2+\xi}{3}})
      \right.
  \nonumber\\
  && \left.
     -\frac{3}{2}(
      \psi^{(-2)}({\scriptstyle\frac{2}{3}})+
      \psi^{(-2)}({\scriptstyle\frac{1}{3}}))\right]
\,,
\end{eqnarray}
where $\psi^{(-2)}(x)=\int_0^{x}\ln\Gamma(u)\,du$~\cite{A98, CS05}.

\bibliography{biblio}

\begin{thebibliography}{31}%
\makeatletter
\providecommand \@ifxundefined [1]{%
 \@ifx{#1\undefined}
}%
\providecommand \@ifnum [1]{%
 \ifnum #1\expandafter \@firstoftwo
 \else \expandafter \@secondoftwo
 \fi
}%
\providecommand \@ifx [1]{%
 \ifx #1\expandafter \@firstoftwo
 \else \expandafter \@secondoftwo
 \fi
}%
\providecommand \natexlab [1]{#1}%
\providecommand \enquote  [1]{``#1''}%
\providecommand \bibnamefont  [1]{#1}%
\providecommand \bibfnamefont [1]{#1}%
\providecommand \citenamefont [1]{#1}%
\providecommand \href@noop [0]{\@secondoftwo}%
\providecommand \href [0]{\begingroup \@sanitize@url \@href}%
\providecommand \@href[1]{\@@startlink{#1}\@@href}%
\providecommand \@@href[1]{\endgroup#1\@@endlink}%
\providecommand \@sanitize@url [0]{\catcode `\\12\catcode `\$12\catcode
  `\&12\catcode `\#12\catcode `\^12\catcode `\_12\catcode `\%12\relax}%
\providecommand \@@startlink[1]{}%
\providecommand \@@endlink[0]{}%
\providecommand \url  [0]{\begingroup\@sanitize@url \@url }%
\providecommand \@url [1]{\endgroup\@href {#1}{\urlprefix }}%
\providecommand \urlprefix  [0]{URL }%
\providecommand \Eprint [0]{\href }%
\providecommand \doibase [0]{http://dx.doi.org/}%
\providecommand \selectlanguage [0]{\@gobble}%
\providecommand \bibinfo  [0]{\@secondoftwo}%
\providecommand \bibfield  [0]{\@secondoftwo}%
\providecommand \translation [1]{[#1]}%
\providecommand \BibitemOpen [0]{}%
\providecommand \bibitemStop [0]{}%
\providecommand \bibitemNoStop [0]{.\EOS\space}%
\providecommand \EOS [0]{\spacefactor3000\relax}%
\providecommand \BibitemShut  [1]{\csname bibitem#1\endcsname}%
\let\auto@bib@innerbib\@empty
%</preamble>
\bibitem [{\citenamefont {Levin}(2002)}]{Levin02}%
  \BibitemOpen
  \bibfield  {author} {\bibinfo {author} {\bibfnamefont {Y.}~\bibnamefont
  {Levin}},\ }\href@noop {} {\bibfield  {journal} {\bibinfo  {journal} {Rep.
  Prog. Phys.}\ }\textbf {\bibinfo {volume} {65}},\ \bibinfo {pages} {1577}
  (\bibinfo {year} {2002})}\BibitemShut {NoStop}%
\bibitem [{\citenamefont {Andelman}(2006)}]{Andelman06}%
  \BibitemOpen
  \bibfield  {author} {\bibinfo {author} {\bibfnamefont {D.}~\bibnamefont
  {Andelman}},\ }in\ \href@noop {} {\emph {\bibinfo {booktitle} {Soft Condensed
  Matter Physics in Molecular and Cell Biology}}},\ \bibinfo {editor} {edited
  by\ \bibinfo {editor} {\bibfnamefont {W.}~\bibnamefont {Poon}}\ and\ \bibinfo
  {editor} {\bibfnamefont {D.}~\bibnamefont {Andelman}}}\ (\bibinfo
  {publisher} {Addison Wesley},\ \bibinfo {year} {2006})\ Chap.~\bibinfo
  {chapter} {6}\BibitemShut {NoStop}%
\bibitem [{\citenamefont {Messina}(2009)}]{Messina09}%
  \BibitemOpen
  \bibfield  {author} {\bibinfo {author} {\bibfnamefont {R.}~\bibnamefont
  {Messina}},\ }\href@noop {} {\bibfield  {journal} {\bibinfo  {journal} {J.
  Phys.: Condens. Matter}\ }\textbf {\bibinfo {volume} {21}},\ \bibinfo {pages}
  {113102} (\bibinfo {year} {2009})}\BibitemShut {NoStop}%
\bibitem [{\citenamefont {Verwey}\ and\ \citenamefont {Overbeek}(1948)}]{VO48}%
  \BibitemOpen
  \bibfield  {author} {\bibinfo {author} {\bibfnamefont {E.~J.~W.}\
  \bibnamefont {Verwey}}\ and\ \bibinfo {author} {\bibfnamefont {J.~T.~G.}\
  \bibnamefont {Overbeek}},\ }\href@noop {} {\emph {\bibinfo {title} {Theory of
  the stability of lyophobic colloids}}}\ (\bibinfo  {publisher} {Elsevier},\
  \bibinfo {year} {1948})\BibitemShut {NoStop}%
\bibitem [{\citenamefont {Chan}\ and\ \citenamefont {Mitchell}(1983)}]{CM83}%
  \BibitemOpen
  \bibfield  {author} {\bibinfo {author} {\bibfnamefont {D.~Y.}\ \bibnamefont
  {Chan}}\ and\ \bibinfo {author} {\bibfnamefont {D.}~\bibnamefont
  {Mitchell}},\ }\href {\doibase https://doi.org/10.1016/0021-9797(83)90087-5}
  {\bibfield  {journal} {\bibinfo  {journal} {Journal of Colloid and Interface
  Science}\ }\textbf {\bibinfo {volume} {95}},\ \bibinfo {pages} {193 }
  (\bibinfo {year} {1983})}\BibitemShut {NoStop}%
\bibitem [{\citenamefont {Stigter}\ and\ \citenamefont {Dill}(1989)}]{SD89}%
  \BibitemOpen
  \bibfield  {author} {\bibinfo {author} {\bibfnamefont {D.}~\bibnamefont
  {Stigter}}\ and\ \bibinfo {author} {\bibfnamefont {K.~A.}\ \bibnamefont
  {Dill}},\ }\href {\doibase 10.1021/j100355a033} {\bibfield  {journal}
  {\bibinfo  {journal} {The Journal of Physical Chemistry}\ }\textbf {\bibinfo
  {volume} {93}},\ \bibinfo {pages} {6737} (\bibinfo {year}
  {1989})}\BibitemShut {NoStop}%
\bibitem [{\citenamefont {Shkel}\ \emph {et~al.}(2006)\citenamefont {Shkel},
  \citenamefont {Ballin},\ and\ \citenamefont {Record}}]{SBR06}%
  \BibitemOpen
  \bibfield  {author} {\bibinfo {author} {\bibfnamefont {I.~A.}\ \bibnamefont
  {Shkel}}, \bibinfo {author} {\bibfnamefont {J.~D.}\ \bibnamefont {Ballin}}, \
  and\ \bibinfo {author} {\bibfnamefont {M.~T.}\ \bibnamefont {Record}},\
  }\href {\doibase 10.1021/bi0520434} {\bibfield  {journal} {\bibinfo
  {journal} {Biochemistry}\ }\textbf {\bibinfo {volume} {45}},\ \bibinfo
  {pages} {8411} (\bibinfo {year} {2006})}\BibitemShut {NoStop}%
\bibitem [{\citenamefont {Bret}\ and\ \citenamefont {Zimm}(1984)}]{BZ84}%
  \BibitemOpen
  \bibfield  {author} {\bibinfo {author} {\bibfnamefont {M.~L.}\ \bibnamefont
  {Bret}}\ and\ \bibinfo {author} {\bibfnamefont {B.~H.}\ \bibnamefont
  {Zimm}},\ }\href {\doibase 10.1002/bip.360230209} {\bibfield  {journal}
  {\bibinfo  {journal} {Biopolymers}\ }\textbf {\bibinfo {volume} {23}},\
  \bibinfo {pages} {287} (\bibinfo {year} {1984})}\BibitemShut {NoStop}%
\bibitem [{\citenamefont {Mills}\ \emph {et~al.}(1985)\citenamefont {Mills},
  \citenamefont {Anderson},\ and\ \citenamefont {Record}}]{MAR89}%
  \BibitemOpen
  \bibfield  {author} {\bibinfo {author} {\bibfnamefont {P.}~\bibnamefont
  {Mills}}, \bibinfo {author} {\bibfnamefont {C.~F.}\ \bibnamefont {Anderson}},
  \ and\ \bibinfo {author} {\bibfnamefont {M.~T.}\ \bibnamefont {Record}},\
  }\href {\doibase 10.1021/j100265a012} {\bibfield  {journal} {\bibinfo
  {journal} {The Journal of Physical Chemistry}\ }\textbf {\bibinfo {volume}
  {89}},\ \bibinfo {pages} {3984} (\bibinfo {year} {1985})}\BibitemShut
  {NoStop}%
\bibitem [{\citenamefont {Trizac}\ \emph {et~al.}(2003)\citenamefont {Trizac},
  \citenamefont {Bocquet}, \citenamefont {Aubouy},\ and\ \citenamefont {von
  Gr\"unberg}}]{TBAG03}%
  \BibitemOpen
  \bibfield  {author} {\bibinfo {author} {\bibfnamefont {E.}~\bibnamefont
  {Trizac}}, \bibinfo {author} {\bibfnamefont {L.}~\bibnamefont {Bocquet}},
  \bibinfo {author} {\bibfnamefont {M.}~\bibnamefont {Aubouy}}, \ and\ \bibinfo
  {author} {\bibfnamefont {H.~H.}\ \bibnamefont {von Gr\"unberg}},\ }\href
  {\doibase 10.1021/la027056m} {\bibfield  {journal} {\bibinfo  {journal}
  {Langmuir}\ }\textbf {\bibinfo {volume} {19}},\ \bibinfo {pages} {4027}
  (\bibinfo {year} {2003})}\BibitemShut {NoStop}%
\bibitem [{\citenamefont {Shkel}(2010)}]{S10}%
  \BibitemOpen
  \bibfield  {author} {\bibinfo {author} {\bibfnamefont {I.~A.}\ \bibnamefont
  {Shkel}},\ }\href {\doibase 10.1021/jp908267c} {\bibfield  {journal}
  {\bibinfo  {journal} {The Journal of Physical Chemistry B}\ }\textbf
  {\bibinfo {volume} {114}},\ \bibinfo {pages} {10793} (\bibinfo {year}
  {2010})}\BibitemShut {NoStop}%
\bibitem [{\citenamefont {McCoy}\ \emph {et~al.}(1977)\citenamefont {McCoy},
  \citenamefont {Tracy},\ and\ \citenamefont {Wu}}]{CTW77}%
  \BibitemOpen
  \bibfield  {author} {\bibinfo {author} {\bibfnamefont {B.~M.}\ \bibnamefont
  {McCoy}}, \bibinfo {author} {\bibfnamefont {C.~A.}\ \bibnamefont {Tracy}}, \
  and\ \bibinfo {author} {\bibfnamefont {T.~T.}\ \bibnamefont {Wu}},\
  }\href@noop {} {\bibfield  {journal} {\bibinfo  {journal} {J. Math. Phys.}\
  }\textbf {\bibinfo {volume} {18}},\ \bibinfo {pages} {1058} (\bibinfo {year}
  {1977})}\BibitemShut {NoStop}%
\bibitem [{\citenamefont {Tracy}\ and\ \citenamefont {Widom}(1997)}]{TW97}%
  \BibitemOpen
  \bibfield  {author} {\bibinfo {author} {\bibfnamefont {C.~A.}\ \bibnamefont
  {Tracy}}\ and\ \bibinfo {author} {\bibfnamefont {H.}~\bibnamefont {Widom}},\
  }\href {\doibase https://doi.org/10.1016/S0378-4371(97)00229-X} {\bibfield
  {journal} {\bibinfo  {journal} {Physica A: Statistical Mechanics and its
  Applications}\ }\textbf {\bibinfo {volume} {244}},\ \bibinfo {pages} {402 }
  (\bibinfo {year} {1997})}\BibitemShut {NoStop}%
\bibitem [{\citenamefont {Tracy}\ and\ \citenamefont {Widom}(1998)}]{TW98}%
  \BibitemOpen
  \bibfield  {author} {\bibinfo {author} {\bibfnamefont {C.~A.}\ \bibnamefont
  {Tracy}}\ and\ \bibinfo {author} {\bibfnamefont {H.}~\bibnamefont {Widom}},\
  }\href@noop {} {\bibfield  {journal} {\bibinfo  {journal} {Commun. Math.
  Phys.}\ }\textbf {\bibinfo {volume} {190}},\ \bibinfo {pages} {697} (\bibinfo
  {year} {1998})}\BibitemShut {NoStop}%
\bibitem [{\citenamefont {Trizac}\ and\ \citenamefont {T\'ellez}(2006)}]{TT06}%
  \BibitemOpen
  \bibfield  {author} {\bibinfo {author} {\bibfnamefont {E.}~\bibnamefont
  {Trizac}}\ and\ \bibinfo {author} {\bibfnamefont {G.}~\bibnamefont
  {T\'ellez}},\ }\href {\doibase 10.1103/PhysRevLett.96.038302} {\bibfield
  {journal} {\bibinfo  {journal} {Phys. Rev. Lett.}\ }\textbf {\bibinfo
  {volume} {96}},\ \bibinfo {pages} {038302} (\bibinfo {year}
  {2006})}\BibitemShut {NoStop}%
\bibitem [{\citenamefont {T{\'{e}}llez}\ and\ \citenamefont
  {Trizac}(2006)}]{TTexactPB06}%
  \BibitemOpen
  \bibfield  {author} {\bibinfo {author} {\bibfnamefont {G.}~\bibnamefont
  {T{\'{e}}llez}}\ and\ \bibinfo {author} {\bibfnamefont {E.}~\bibnamefont
  {Trizac}},\ }\href {\doibase 10.1088/1742-5468/2006/06/p06018} {\bibfield
  {journal} {\bibinfo  {journal} {Journal of Statistical Mechanics: Theory and
  Experiment}\ }\textbf {\bibinfo {volume} {2006}},\ \bibinfo {pages} {P06018}
  (\bibinfo {year} {2006})}\BibitemShut {NoStop}%
\bibitem [{\citenamefont {Trizac}\ and\ \citenamefont {T\'ellez}(2007)}]{TT07}%
  \BibitemOpen
  \bibfield  {author} {\bibinfo {author} {\bibfnamefont {E.}~\bibnamefont
  {Trizac}}\ and\ \bibinfo {author} {\bibfnamefont {G.}~\bibnamefont
  {T\'ellez}},\ }\href {\doibase 10.1021/ma061497l} {\bibfield  {journal}
  {\bibinfo  {journal} {Macromolecules}\ }\textbf {\bibinfo {volume} {40}},\
  \bibinfo {pages} {1305} (\bibinfo {year} {2007})}\BibitemShut {NoStop}%
\bibitem [{\citenamefont {Marcus}(1955)}]{M55}%
  \BibitemOpen
  \bibfield  {author} {\bibinfo {author} {\bibfnamefont {R.~A.}\ \bibnamefont
  {Marcus}},\ }\href@noop {} {\bibfield  {journal} {\bibinfo  {journal} {The
  Journal of Chemical Physics}\ }\textbf {\bibinfo {volume} {23}},\ \bibinfo
  {pages} {1057} (\bibinfo {year} {1955})}\BibitemShut {NoStop}%
\bibitem [{\citenamefont {Trizac}\ and\ \citenamefont {Hansen}(1996)}]{TH96}%
  \BibitemOpen
  \bibfield  {author} {\bibinfo {author} {\bibfnamefont {E.}~\bibnamefont
  {Trizac}}\ and\ \bibinfo {author} {\bibfnamefont {J.-P.}\ \bibnamefont
  {Hansen}},\ }\href {\doibase 10.1088/0953-8984/8/47/008} {\bibfield
  {journal} {\bibinfo  {journal} {Journal of Physics: Condensed Matter}\
  }\textbf {\bibinfo {volume} {8}},\ \bibinfo {pages} {9191} (\bibinfo {year}
  {1996})}\BibitemShut {NoStop}%
\bibitem [{\citenamefont {Trizac}\ and\ \citenamefont {Hansen}(1997)}]{TH97}%
  \BibitemOpen
  \bibfield  {author} {\bibinfo {author} {\bibfnamefont {E.}~\bibnamefont
  {Trizac}}\ and\ \bibinfo {author} {\bibfnamefont {J.-P.}\ \bibnamefont
  {Hansen}},\ }\href {\doibase 10.1103/PhysRevE.56.3137} {\bibfield  {journal}
  {\bibinfo  {journal} {Phys. Rev. E}\ }\textbf {\bibinfo {volume} {56}},\
  \bibinfo {pages} {3137} (\bibinfo {year} {1997})}\BibitemShut {NoStop}%
\bibitem [{\citenamefont {Fuoss}\ \emph {et~al.}(1951)\citenamefont {Fuoss},
  \citenamefont {Katchalsky},\ and\ \citenamefont {Lifson}}]{FKL51}%
  \BibitemOpen
  \bibfield  {author} {\bibinfo {author} {\bibfnamefont {R.~M.}\ \bibnamefont
  {Fuoss}}, \bibinfo {author} {\bibfnamefont {A.}~\bibnamefont {Katchalsky}}, \
  and\ \bibinfo {author} {\bibfnamefont {S.}~\bibnamefont {Lifson}},\
  }\href@noop {} {\bibfield  {journal} {\bibinfo  {journal} {Proc Natl Acad Sci
  U S A}\ }\textbf {\bibinfo {volume} {37}},\ \bibinfo {pages} {579} (\bibinfo
  {year} {1951})}\BibitemShut {NoStop}%
\bibitem [{\citenamefont {Alfrey~Jr.}\ \emph {et~al.}(1951)\citenamefont
  {Alfrey~Jr.}, \citenamefont {Berg},\ and\ \citenamefont {Morawetz}}]{ABM51}%
  \BibitemOpen
  \bibfield  {author} {\bibinfo {author} {\bibfnamefont {T.}~\bibnamefont
  {Alfrey~Jr.}}, \bibinfo {author} {\bibfnamefont {P.~W.}\ \bibnamefont
  {Berg}}, \ and\ \bibinfo {author} {\bibfnamefont {H.}~\bibnamefont
  {Morawetz}},\ }\href@noop {} {\bibfield  {journal} {\bibinfo  {journal}
  {Journal of Polymer Science}\ }\textbf {\bibinfo {volume} {7}},\ \bibinfo
  {pages} {543} (\bibinfo {year} {1951})}\BibitemShut {NoStop}%
\bibitem [{\citenamefont {Netz}(2001)}]{Netz01}%
  \BibitemOpen
  \bibfield  {author} {\bibinfo {author} {\bibfnamefont {R.}~\bibnamefont
  {Netz}},\ }\href@noop {} {\bibfield  {journal} {\bibinfo  {journal} {Eur.
  Phys. J. E}\ }\textbf {\bibinfo {volume} {5}},\ \bibinfo {pages} {557}
  (\bibinfo {year} {2001})}\BibitemShut {NoStop}%
\bibitem [{\citenamefont {Alexander}\ \emph {et~al.}(1984)\citenamefont
  {Alexander}, \citenamefont {Chaikin}, \citenamefont {Grant}, \citenamefont
  {Morales}, \citenamefont {Pincus},\ and\ \citenamefont {Hone}}]{ACGMPH84}%
  \BibitemOpen
  \bibfield  {author} {\bibinfo {author} {\bibfnamefont {S.}~\bibnamefont
  {Alexander}}, \bibinfo {author} {\bibfnamefont {P.~M.}\ \bibnamefont
  {Chaikin}}, \bibinfo {author} {\bibfnamefont {P.}~\bibnamefont {Grant}},
  \bibinfo {author} {\bibfnamefont {G.~J.}\ \bibnamefont {Morales}}, \bibinfo
  {author} {\bibfnamefont {P.}~\bibnamefont {Pincus}}, \ and\ \bibinfo {author}
  {\bibfnamefont {D.}~\bibnamefont {Hone}},\ }\href {\doibase 10.1063/1.446600}
  {\bibfield  {journal} {\bibinfo  {journal} {The Journal of Chemical Physics}\
  }\textbf {\bibinfo {volume} {80}},\ \bibinfo {pages} {5776} (\bibinfo {year}
  {1984})}\BibitemShut {NoStop}%
\bibitem [{\citenamefont {Aubouy}\ \emph {et~al.}(2003)\citenamefont {Aubouy},
  \citenamefont {Trizac},\ and\ \citenamefont {Bocquet}}]{ATB03}%
  \BibitemOpen
  \bibfield  {author} {\bibinfo {author} {\bibfnamefont {M.}~\bibnamefont
  {Aubouy}}, \bibinfo {author} {\bibfnamefont {E.}~\bibnamefont {Trizac}}, \
  and\ \bibinfo {author} {\bibfnamefont {L.}~\bibnamefont {Bocquet}},\ }\href
  {\doibase 10.1088/0305-4470/36/22/302} {\bibfield  {journal} {\bibinfo
  {journal} {Journal of Physics A: Mathematical and General}\ }\textbf
  {\bibinfo {volume} {36}},\ \bibinfo {pages} {5835} (\bibinfo {year}
  {2003})}\BibitemShut {NoStop}%
\bibitem [{\citenamefont {Trizac}\ and\ \citenamefont {Shen}(2016)}]{ShTr16}%
  \BibitemOpen
  \bibfield  {author} {\bibinfo {author} {\bibfnamefont {E.}~\bibnamefont
  {Trizac}}\ and\ \bibinfo {author} {\bibfnamefont {T.}~\bibnamefont {Shen}},\
  }\href@noop {} {\bibfield  {journal} {\bibinfo  {journal} {{EPL} (Europhysics
  Letters)}\ }\textbf {\bibinfo {volume} {116}},\ \bibinfo {pages} {18007}
  (\bibinfo {year} {2016})}\BibitemShut {NoStop}%
\bibitem [{\citenamefont {T\'ellez}\ and\ \citenamefont {Trizac}(2003)}]{TT03}%
  \BibitemOpen
  \bibfield  {author} {\bibinfo {author} {\bibfnamefont {G.}~\bibnamefont
  {T\'ellez}}\ and\ \bibinfo {author} {\bibfnamefont {E.}~\bibnamefont
  {Trizac}},\ }\href {\doibase 10.1103/PhysRevE.68.061401} {\bibfield
  {journal} {\bibinfo  {journal} {Phys. Rev. E}\ }\textbf {\bibinfo {volume}
  {68}},\ \bibinfo {pages} {061401} (\bibinfo {year} {2003})}\BibitemShut
  {NoStop}%
\bibitem [{\citenamefont {Shkel}\ \emph {et~al.}(2002)\citenamefont {Shkel},
  \citenamefont {Tsodikov},\ and\ \citenamefont {Record}}]{STR02}%
  \BibitemOpen
  \bibfield  {author} {\bibinfo {author} {\bibfnamefont {I.~A.}\ \bibnamefont
  {Shkel}}, \bibinfo {author} {\bibfnamefont {O.~V.}\ \bibnamefont {Tsodikov}},
  \ and\ \bibinfo {author} {\bibfnamefont {J.}~\bibnamefont {Record},
  \bibfnamefont {M.~T.}},\ }\href@noop {} {\bibfield  {journal} {\bibinfo
  {journal} {Proc. Natl. Acad. Sci. U.S.A.}\ }\textbf {\bibinfo {volume}
  {99}},\ \bibinfo {pages} {2597} (\bibinfo {year} {2002})}\BibitemShut
  {NoStop}%
\bibitem [{Note1()}]{Note1}%
  \BibitemOpen
  \bibinfo {note} {In Ref.~\cite {TT07} the cases 1:2 and 2:1 were inverted in
  Eq. (D2).}\BibitemShut {Stop}%
\bibitem [{\citenamefont {Adamchik}(1998)}]{A98}%
  \BibitemOpen
  \bibfield  {author} {\bibinfo {author} {\bibfnamefont {V.~S.}\ \bibnamefont
  {Adamchik}},\ }\href {\doibase https://doi.org/10.1016/S0377-0427(98)00192-7}
  {\bibfield  {journal} {\bibinfo  {journal} {Journal of Computational and
  Applied Mathematics}\ }\textbf {\bibinfo {volume} {100}},\ \bibinfo {pages}
  {191 } (\bibinfo {year} {1998})}\BibitemShut {NoStop}%
\bibitem [{\citenamefont {Choi}\ and\ \citenamefont {Srivastava}(2005)}]{CS05}%
  \BibitemOpen
  \bibfield  {author} {\bibinfo {author} {\bibfnamefont {J.}~\bibnamefont
  {Choi}}\ and\ \bibinfo {author} {\bibfnamefont {H.}~\bibnamefont
  {Srivastava}},\ }\href {\doibase https://doi.org/10.1016/j.jmaa.2004.08.043}
  {\bibfield  {journal} {\bibinfo  {journal} {Journal of Mathematical Analysis
  and Applications}\ }\textbf {\bibinfo {volume} {303}},\ \bibinfo {pages} {436
  } (\bibinfo {year} {2005})}\BibitemShut {NoStop}%
\end{thebibliography}%

\end{document}